\documentclass[pdftex,citeautoscript,aps,prl,superscriptaddress,cite,amsmath,amsfonts,amssymb,twocolumn]{revtex4}
\pdfoutput=1
\renewcommand{\vec}[1]{\mathbf{#1}} 
\usepackage[pdftex]{graphicx}
\usepackage{color}

\newcommand{\figref}[1]{Fig.~\ref{fig:#1}}

\newcommand{\eqnumref}[1]{(\ref{eq:#1})}
\renewcommand{\eqref}[1]{Eq.~\eqnumref{#1}}

\newcommand{\citeasnoun}[1]{Ref.~\onlinecite{#1}}

\begin{document}
\title{Microstructure Effects for Casimir Forces in Chiral Metamaterials}

\author{Alexander~P.~McCauley}
\affiliation{Department of Physics, Massachusetts Institute of Technology, Cambridge MA 02139, USA}
\author{Rongkuo~Zhao}
\affiliation{Ames Laboratory and Department of Physics and Astronomy, Iowa State University, Ames, Iowa 50011, USA}
\author{M.~T.~Homer~Reid}
\affiliation{Department of Physics, Massachusetts Institute of Technology, Cambridge MA 02139, USA}
\author{Alejandro~W.~Rodriguez}
\affiliation{Department of Physics, Massachusetts Institute of Technology, Cambridge MA 02139, USA}
\author{Jiangfeng~Zhou}
\affiliation{MPA-CINT, MS K771, Los Alamos National Laboratory, Los Alamos, New Mexico 87545, USA}
\author{F.~S.~S.~Rosa}
\affiliation{Theoretical Division MS B213, Los Alamos National Laboratory, Los Alamos, New Mexico 87545, USA}
\author{John~D.~Joannopoulos}
\affiliation{Department of Physics, Massachusetts Institute of Technology, Cambridge MA 02139, USA}
\author{D.~A.~R.~Dalvit}
\affiliation{Theoretical Division MS B213, Los Alamos National Laboratory, Los Alamos, New Mexico 87545, USA}
\author{Costas~M.~Soukoulis}
\affiliation{Ames Laboratory and Department of Physics and Astronomy, Iowa State University, Ames, Iowa 50011, USA}
\author{Steven~G.~Johnson}
\affiliation{Department of Mathematics, Massachusetts Institute of Technology, Cambridge MA 02139, USA}

\date{\today}

\begin{abstract}
We examine a recent prediction for the chirality-dependence of the
Casimir force in chiral metamaterials by numerical computation of the
forces between the exact microstructures, rather than homogeneous
approximations.  We compute the exact force for a chiral bent-cross
pattern, as well as forces for an idealized ``omega''-particle medium
in the dilute approximation and identify the effects of structural
inhomogeneity (i.e. proximity forces and anisotropy).  We find that
these microstructure effects dominate the force for separations where
chirality was predicted to have a strong influence.  To get
observations of chirality free from microstructure effects, one must
go to large separations where the effect of chirality is at most
$\sim10^{-4}$ of the total force.
\end{abstract}

\maketitle

It has been proposed that dielectric metamaterials might exhibit
repulsive Casimir forces in vacuum where planar structures have only
attraction~\cite{Kenneth02,Henkel05,Leonhardt07,Rosa08:PRL}.  However,
these predictions used effective-medium approximations (EMAs) for the
metamaterials, often treating the EMA terms as free parameters.  While
certain parameters give repulsion, a recent theorem~\cite{Rahi10:PRL}
implies that one cannot have repulsion between vacuum-separated
dielectric/metallic metamaterials in the effective--medium (large
separation) regime (although it is possible outside of this
regime~\cite{LevinMc10}).  However, it is possible that significant
reduction or modulation of the Casimir force can occur as a result of
metamaterial effects.  A recent EMA analysis of chiral
metamaterials~\cite{Zhao09} predicted such a force reduction, and even
repulsion, due to chirality.  Unfortunately, chirality effects vanish
at large separations where the EMA should be valid, and are strongest
at small separations where the EMA is questionable.  Therefore, the
question of whether chiral metamaterial effects can theoretically
modulate the Casimir force remains open.  Answering it requires
accurate calculations using the exact microstructure of the
metamaterials, rather than a homogeneous EMA.  Recent
advances~\cite{Emig07, Kenneth08, Neto08, Rahi09:PRD, ReidRo09,
  RodriguezMc09:PRA, McCauleyRo10:PRA} have made the treatment of such
complex structures possible.  In this paper, we apply these methods to
rigorously test the EMA predictions for chiral metamaterials against
numerical calculations incorporating the microstructures.  We are able
to distinguish chiral metamaterial effects from other ``non-ideal''
effects such as pairwise surface-surface attractions, and find that
the former are overwhelmed except at large separations, where they are
only $10^{-4}$ or less of the total force.


Dielectric and metallic metamaterials are defined by an inhomogeneous
permittivity $\varepsilon(\vec{x},\omega)$, but at sufficiently long
wavelengths (large separations in the Casimir context) they can be
accurately modeled in the EMA by homogeneous constitutive parameters
[e.g, $\varepsilon(\omega)$, $\mu(\omega)$] that can be very different
from the constituent materials.  The Casimir force between two bodies,
however, is naturally expressed as an integral over imaginary
frequency $\omega = i\xi$~\cite{Landau:stat2}.  The force in the EMA
will then depend on the effective $\varepsilon(i\xi)$, $\mu(i\xi)$,
etc.\ over a range of imaginary frequencies~\cite{Rosa08:PRL}.  Early
works~\cite{Kenneth02,Henkel05,Leonhardt07} predicted repulsive effects by a
putative $\varepsilon(i\xi) < \mu(i\xi)$ for one body.  However, no
such repulsion was found for ``magnetic'' metamaterials based on
actual structures, in which $\varepsilon(i\xi) > \mu(i\xi)$ for all
$\xi$~\cite{Rosa08:PRL, Rosa09}.  Basically, while these metamaterials
can have almost any $\varepsilon$ and $\mu$ at a given \emph{real}
$\omega$ via resonances, on the imaginary-$\omega$ axis the important
features come from the behavior at low $\xi$ (long wavelengths).  In
this limit, $\mu(i\xi)- \mu(0) \sim -\xi^2$~\cite{Tretyakov2005,
  Rosa08:PRL, Rosa09}, where $\mu(0) < \varepsilon(0)$.  Therefore,
there is no repulsion or force reduction in the EMA regime for this
model.  More recently,~\citeasnoun{Zhao09} studied chiral
metamaterials, which in addition to $\varepsilon$ and $\mu$ are
characterized in the EMA by a chirality $\kappa$ coupling $\vec{D}$ to
$\vec{H}$ and $\vec{B}$ to $\vec{E}$.  $\kappa \rightarrow -\kappa$ by
a spatial inversion ($\vec{x}\rightarrow -\vec{x}$) of the
microstructure.  For any $\kappa$, the EMA predicts that the force
between two media of the same chirality (SC) should be lower than for
media of the opposite chirality (OC), with the size of this difference
increasing with $\kappa$.  Because in general $\kappa\sim
\xi$~\cite{Tretyakov2005, Zhao09}, this effect is largest at short
separations and goes to zero at large separations.  Therefore, to get
a useful prediction (e.g., force reduction due to chirality), we are
forced to consider the predictions of the EMA at intermediate
separations.  However, not only does the EMA fail at sufficiently
small separations, but the force in that limit is eventually described
by the proximity force approximation (PFA)~\cite{Derjaguin56} in which
there are only pairwise attractions.  One should therefore be cautious
of any EMA prediction that differs qualitatively from PFA in this
limit.  An analysis based on pairwise attractions would find little
effect due to chirality.  Instead, the behavior would be dominated by
small-scale features such as the shortest distance between two
components of the microstructures.  Thus far, no attempt has been made
to determine which behavior dominates, and we do this as follows: if
the materials behave as homogeneous, chiral media, the relative
chirality should be the only source of a force difference between the
SC and OC cases, where $F_{OC}> F_{SC}$, independent of the transverse
displacement~$x$.  However, if the force is governed by pairwise
attraction, it should exhibit a strong $x$-dependence. We can
therefore directly test the validity of the EMA by comparing the force
for different values of~$x$.


\begin{figure}[tb]
\includegraphics[width=0.9\columnwidth]{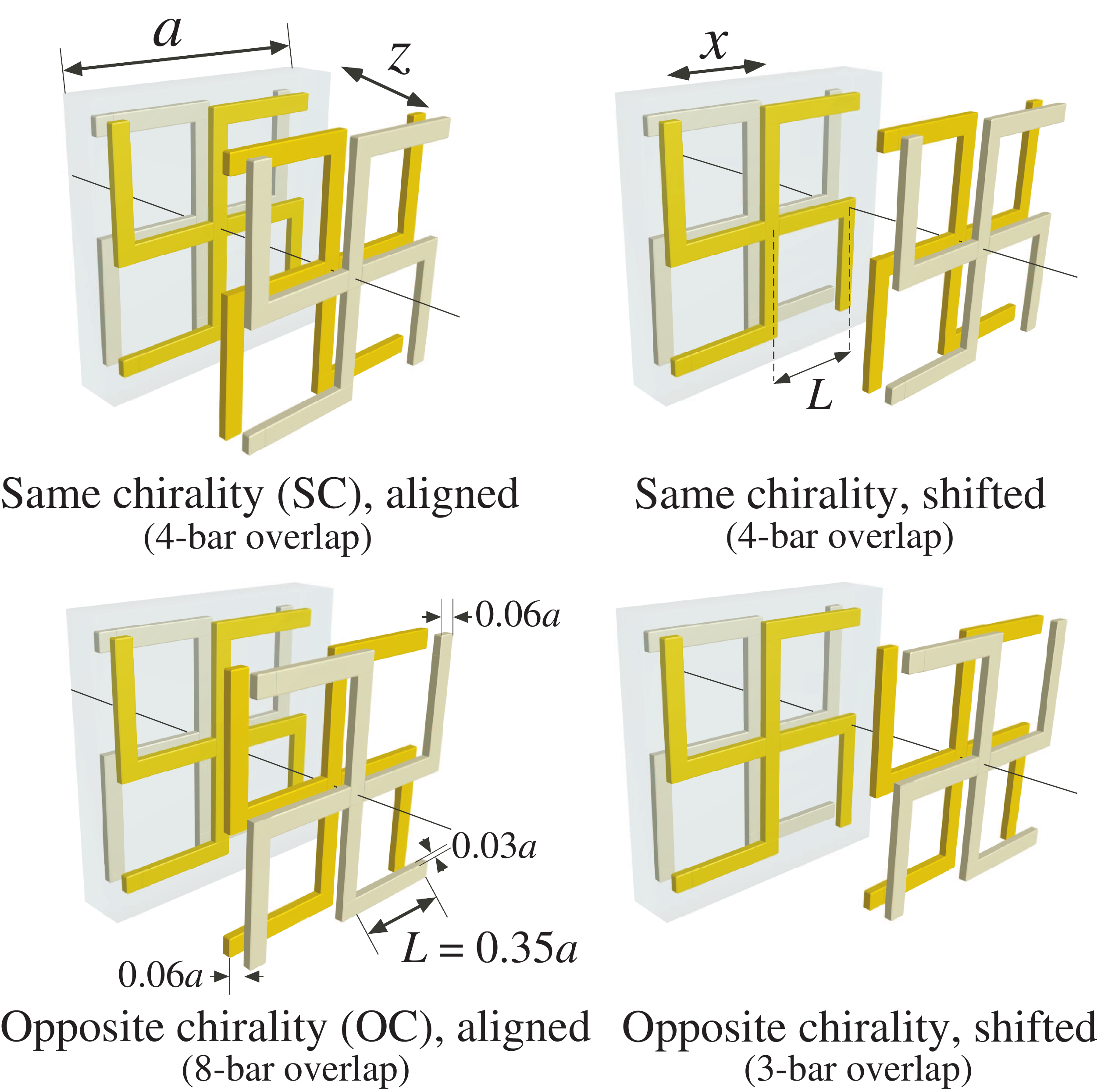}
\centering
\caption{(color online) Interactions between two chiral
  metamaterials~\cite{Zhao10:PRB}.  Shaded box indicates a unit
  cell; the structures are periodic in the transverse direction.
  Nearest neighbor bars on opposing structures are colored yellow.  In
  the EMA, the forces should obey $F_{SC} < F_{OC}$.  By contrast, in
  the pairwise-force approximation the force is determined by the
  number of overlapping nearest-neighbor bars, as indicated: $F_{OC} >
  F_{SC}$ when the centers are aligned (left column), but $F_{OC} <
  F_{SC}$ when they are displaced by $L/2$ (right column).}
\label{fig:cw-configs}
\end{figure}

\begin{figure}[tb]
\includegraphics[width=1.0\columnwidth]{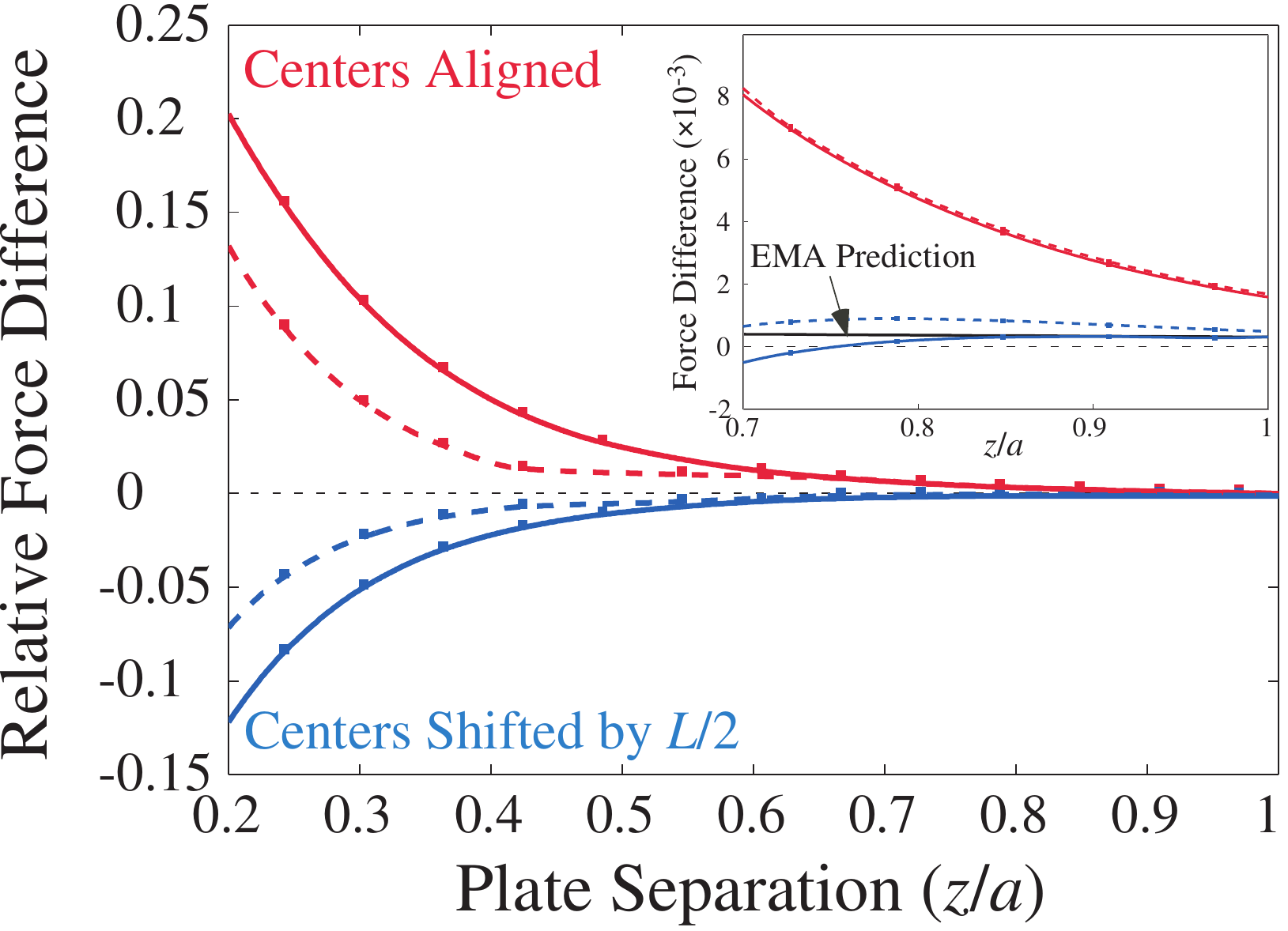}
\centering
\caption{(color online) Relative force difference
  $(F_{OC}-F_{SC})/F_{OC}$ for the left (red) and right (blue) columns
  of~\figref{cw-configs} for PEC (solid) and dispersive gold for $a =
  1~\mu m$ (dashed).  From~\figref{cw-configs}, the sign difference is
  due to proximity effects.  The inset shows a zoom of the large-$z$
  regime, along with the EMA prediction (black).}
\label{fig:cw-force}
\end{figure}

We first examine a realistic structure proposed
in~\citeasnoun{Zhao10:PRB} and shown in~\figref{cw-configs}.  A unit
cell of each ``medium'' consists of a single bi-layer of two bent-arm
crosses, one spiraling clockwise and the other counter-clockwise.
Their ordering in the $z$-direction determines the chirality of the
medium.  We omit the dielectric polyimide in which the metal was
embedded in~\citeasnoun{Zhao10:PRB}, as this eliminates a
chirality-independent attraction between the layers.  The exact
Casimir force between the periodic structures of~\figref{cw-configs}
is computed using a finite-difference time-domain (FDTD)
method~\cite{RodriguezMc09:PRA, McCauleyRo10:PRA}.  Ten bi-layers in
the $z$-direction are included on each side; adding more layers does
not change the results.  We compute the force for two different
material types: perfect electric conductors (PEC) and dispersive gold.
For gold we take a plasma model with $\omega_p = 1.37 \times 10^{16}
\mathrm{rad/sec}$.  The latter model requires a definite value for the
length scale $a$, which we take to be $a = 1 \mu m$.  The results
shown in~\figref{cw-force} (each force difference is normalized by
$F_{OC}(x)$) are similar for both PEC and dispersive gold, and are not
consistent with the EMA: even the sign of $F_{OC}(x) - F_{SC}(x)$ can
be changed as a function of $x$ for $z/a < 0.75$.  Larger $z/a$ still
exhibit a strong $x$-dependence in the force.

These results for low $z/a$ can be qualitatively described by pairwise
nearest-neighbor attractions.  Consider, for instance, the two cases
diagrammed in~\figref{cw-configs}.  First, when the centers of the
unit cells are aligned, there is approximately a 4-bar overlap in the
SC case and an 8-bar overlap in the OC case.  If the force is
proportional to the number of overlapping nearest-neighbor bars, we
expect $F_{OC} > F_{SC}$.  When the centers are displaced by $x=L/2$,
SC has a 4-bar overlap and OC has a 3-bar overlap, so we expect
$F_{OC} < F_{SC}$, and the relative force difference should be reduced
by approximately $1/4$.  This prediction, based purely on pairwise
attraction, captures the behavior for $z/a < 0.75$.  For larger $z/a$
the sign of $F_{OC} - F_{SC}$ is the same in both cases, and the
magnitude of the force difference is comparable to the EMA prediction.
However, in this limit chiral effects are very small (accounting for
only $0.1\%$ of the force).  Furthermore, as the force difference is
still highly $x$-dependent, and other transverse displacements may
still switch the sign of the force difference.  We therefore cannot
determine if this is really an ``ideal'' chiral effect or not.

\begin{figure}[tb]
\includegraphics[width=1.0\columnwidth]{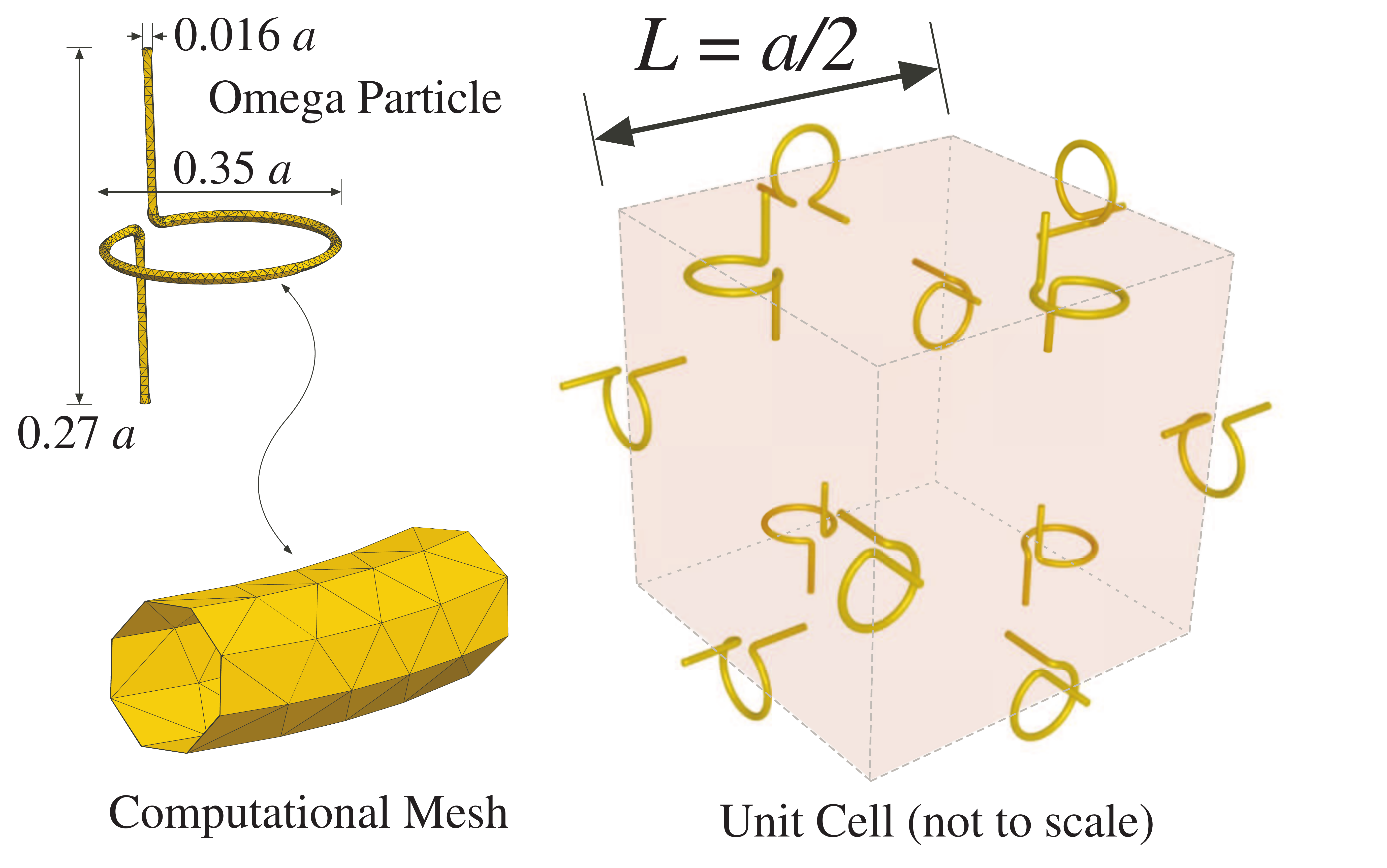}
\includegraphics[width=1.0\columnwidth]{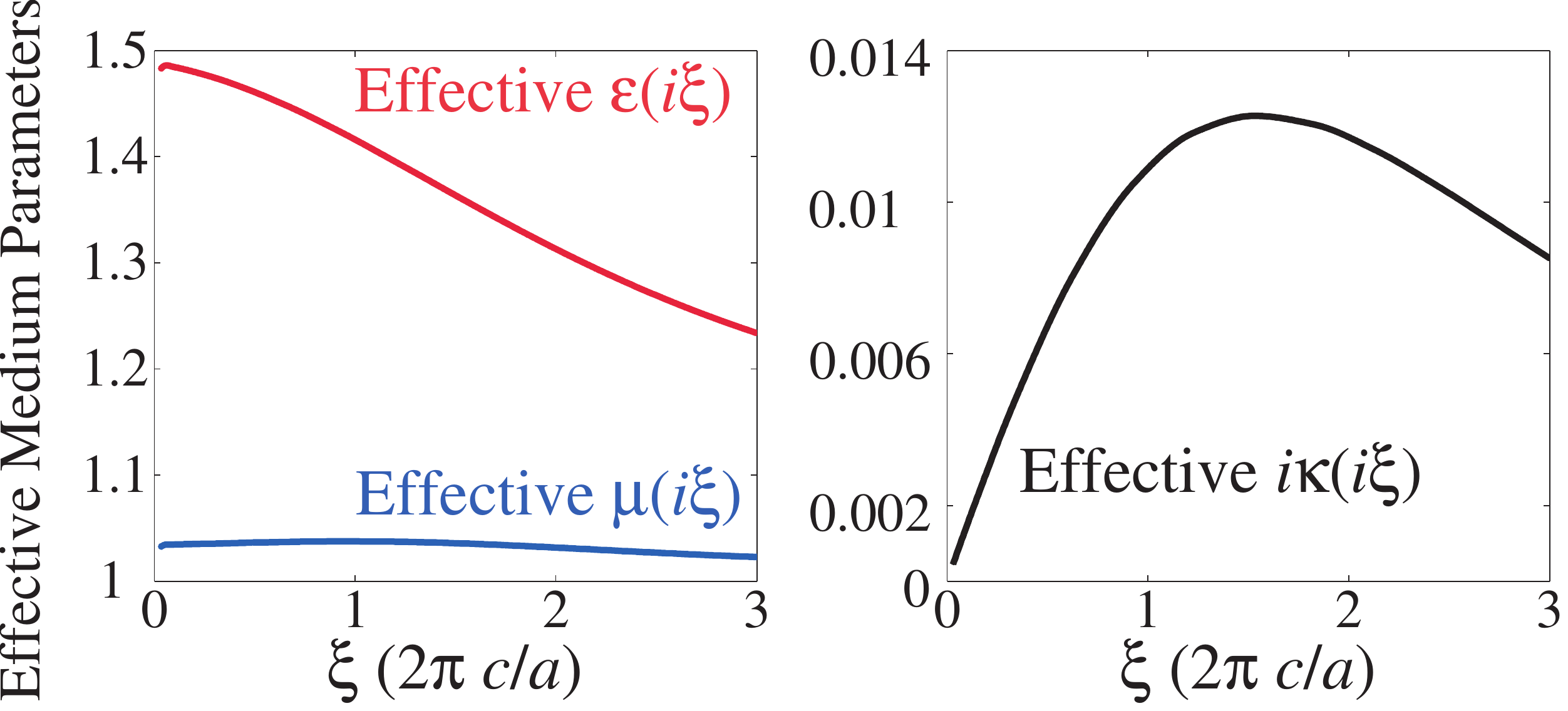}
\centering
\caption{(color online) \emph{Top:} Unit cell of isotropic chiral
  metamaterial: PEC ``omega'' particles in vacuum. Shaded cube
  indicates relative positions in unit cell.  \emph{Bottom:}
  Effective-medium parameters, as a function of imaginary frequency
  $\omega=i\xi$, deduced directly from imaginary-$\omega$ scattering
  data.  $\mu(0) \neq 1$ because the particles are PEC.}
\label{fig:omega-medium}
\end{figure}

As we have seen, when the chirality of the microstructure is thought
to have a large effect on the force, this effect can actually be
attributed to pairwise forces rather than chiral metamaterial effects.
However, the predictions of~\citeasnoun{Zhao10:PRB} were based on
\emph{isotropic}, chiral media, and the bent-arm crosses
of~\figref{cw-configs} are highly anisotropic.  It is possible that
the effects of this anisotropy dominate those of chirality.  To obtain
a chirality with minimal anisotropy, we now consider a more idealized
system: the so-called ``omega''
particles~\cite{Tretyakov2005,Zhao10:PRB}, shown
in~\figref{omega-medium}.  Each individual particle was predicted to
have a strong chiral response, and when assembled into a period-$a$
unit cell of high symmetry should form an isotropic chiral
metamaterial.  The unit cell configuration is shown
in~\figref{omega-medium} (top).  \figref{omega-medium} (bottom) shows
the obtained EMA parameters (discussed below), making this an
isotropic chiral metamaterial.  In fact, by picking a unit cell of
lower symmetry (see below), we find that anisotropy quickly dominates
chiral metamaterial effects.  To make the medium as homogeneous as
possible, the omega particles are relatively closely packed, and as
dispersion had little qualitative role in the previous results, we
take the particles to be PEC for simplicity.

Force computations for a structure as complex as~\figref{omega-medium}
are very difficult for finite-difference methods, because of the
disparity of size between the periodicity $a$ and the wire diameter
$0.016a$.  Instead, we employ a boundary-element
formulation~\cite{ReidRo09} that computes the scattering matrices of
objects in imaginary $\omega$ using a nonuniform mesh of the surfaces,
shown in~\figref{omega-medium}.  In addition, we use a dilute
approximation (justified later) in which multiple scattering events
within a given structure are neglected, so only the scattering
matrices of individual particles are required.  The force between two
periodic structures is then computed from the scattering matrices (see
\citeasnoun{Rahi09:PRD} for a detailed derivation, and a partial
review of precursors~\cite{Emig07, Kenneth08, Neto08}).  The force
computation is summarized as follows: first, the scattering matrix for
each omega particle is numerically computed in a spherical multipole
basis.  This leads to a matrix of scattering amplitudes $F_{l^\prime,
  m^\prime, P^\prime;l,m,P}$, where $l$, $-l\leq m \leq l$ and
$l^\prime$, $-l^\prime \leq m^\prime \leq l^\prime$ are the respective
moments of the incident and scattered spherical waves, and $P,P^\prime
\in \lbrace M, E\rbrace$ their polarization.  Second, the scattering
matrix is converted to a planewave basis, where for each $\omega$ the
planewave has transverse wavevector $\vec{k}_\perp$ and polarization
$P = s,~p$.  Third, to get the scattering from the unit cell
of~\figref{omega-medium} we sum the scattering matrices from the
twelve individual omega particles, each rotated and displaced
appropriately.  The rotations are applied to the spherical multipole
moments via a rotation of the spherical harmonics, and the
displacements are applied in the planewave basis via planewave
translation matrices.  Fourth, the two-dimensional periodicity of the
lattice is incorporated into the problem: for scattering from a unit
cell in the planewave basis, one computes the scattering coefficients
between planewaves of arbitrary $\vec{k}_\perp$ and
$\vec{k}_\perp^\prime$, $P$ and $P^\prime$.  When the scattered fields
are summed over all unit cells of a periodic structure (in two
dimensions), the scattering matrix gains a factor $\sum_\vec{G}
\delta^{(2)}(\vec{k}_\perp - \vec{k}_\perp^\prime + \vec{G}) / a^2$,
where $\lbrace \vec{G} \rbrace$ are the reciprocal lattice vectors.
Therefore, only scattering between planewaves where
$\vec{k}_\perp-\vec{k}_\perp^\prime = \vec{G}$ are needed.  Finally,
to simulate a semi-infinite omega medium in $z$, we apply the
translation matrices to the scattering amplitude of an entire unit
cell, displaced by integer multiples of $a$ in $z$.  With this method,
we can quickly compute forces for many configurations, e.g., many
$x-z$ displacements.  For the present computations, we find that
$l\leq 3$ and $\vec{k}_\perp$ within the first three Brillouin zones
suffice to get the force (and the force difference) to high precision.

To demonstrate that we have an isotropic, chiral medium in the EMA, we
extract the effective-medium parameters $\varepsilon(i\xi)$,
$\mu(i\xi)$, and $\kappa(i\xi)$~[\figref{omega-medium} (bottom)] from
the scattering matrix at $\vec{k}_\perp$.  Parameter
retrieval~\cite{Zhao10:PRB} from reflection and transmission at
normal incidence cannot be used, because for imaginary $\omega$ the
transmission decreases exponentially with the thickness of the medium.
Instead, we compute the specular reflection coefficients
$R_\mathrm{ss}(\omega, \vec{k}_\perp)$ and $R_\mathrm{sp}(\omega,
\vec{k}_\perp)$.  For each frequency $\omega = i\xi$, these quantities
(given in~\citeasnoun{Zhao09}) can be expanded to quadratic order in
$|k_\perp| \ll \xi$ and $\kappa$.  The coefficients of each term are
determined by a fit (with $|\vec{k}_\perp| < \xi/10$).

\begin{figure}[tb]
\includegraphics[width=1.0\columnwidth]{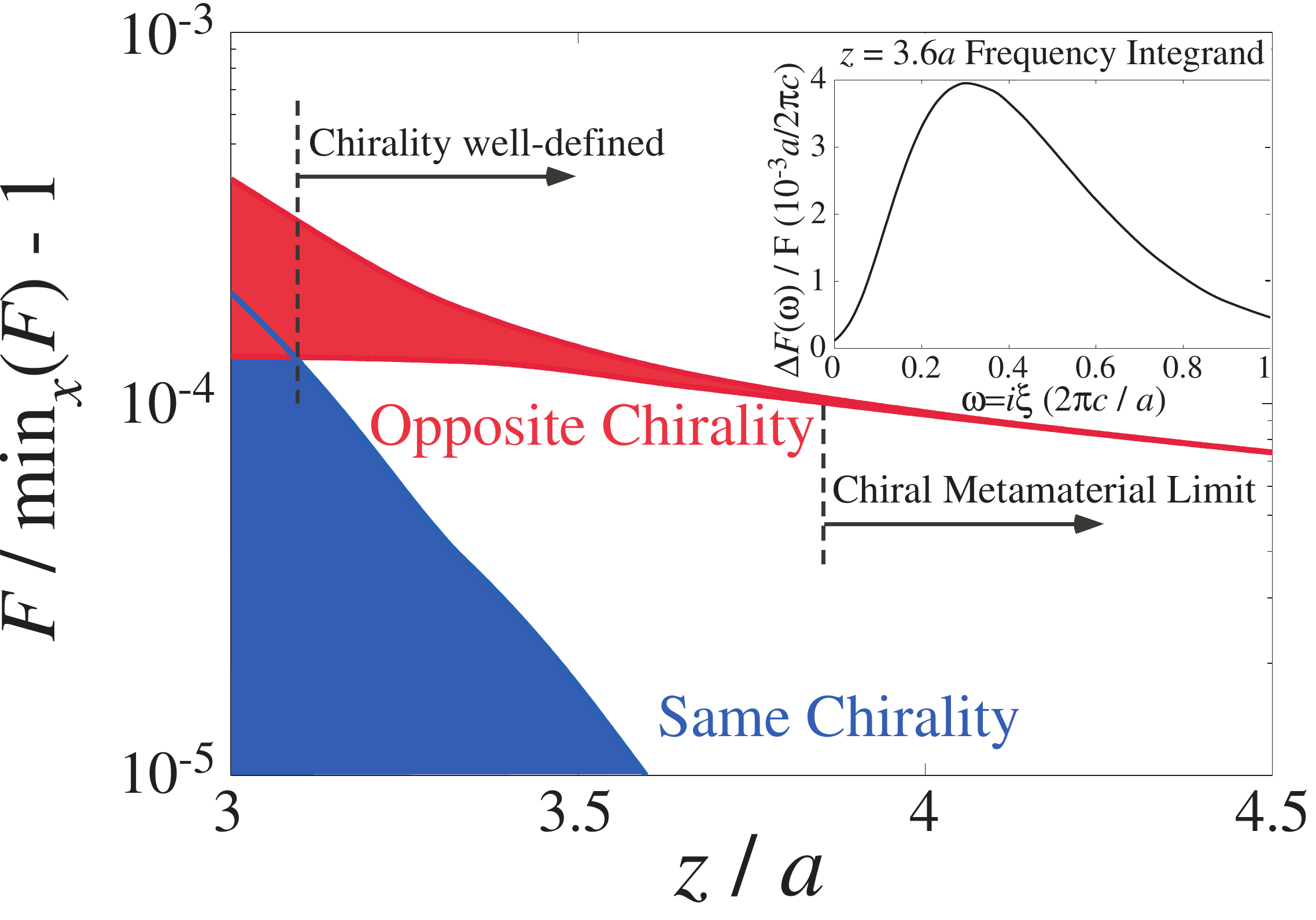}
\centering
\caption{(color online) Forces (in dilute approximation) between
  chiral media from~\figref{omega-medium}, for the SC (blue) and OC
  (red) cases.  The range of the force across all transverse $(x)$
  displacements is shaded, and normalized by the minimum of the force
  over all $x$.  For $z \lesssim 3.1a$, chirality is not well-defined
  (curves overlap).  Only when curves are distinct and $x$-independent
  can the systems be described as chiral metamaterials.  \emph{Inset:}
  Frequency integrand of the relative force difference at $z = 3.6 a$.
  The smallness of the $\omega = 0$ term is an indicator that this
  difference is due to chirality and not anisotropy (see text).}
\label{fig:omega-results}
\end{figure}

The results of the force computations are shown
in~\figref{omega-results} and are similar in form
to~\figref{cw-force}.  We examine the relative force difference at
each $z/a$---normalizing by the minimum force is convenient to obtain
a positive difference that can be plotted on a log scale.  The shaded
areas of each color represent the range of values that the force
assumes for all transverse $(x)$ displacements.  Their spread (the
$x$-dependence) indicates breakdown of the EMA, and for $z \lesssim
3.1 a$ it is so severe that the red and blue force curves overlap.
When these curves separate, but before they sharpen (the blue (SC)
curve goes to zero), we have an intermediate regime where chiral
effects are competing with proximity effects.  Only when the curve
thickness is much less than the relative force difference ($\lesssim
10\%$ for $z \gtrsim 3.6 a$) is EMA accurate.  As mentioned above, we
designed the unit cell to be highly isotropic, for which the EMA
parameters $\varepsilon, \mu, \kappa$ are truly scalars.  For a less
isotropic unit cell (e.g., only one omega particle per unit cell), the
relative force difference can be made much larger, but the source of
this difference is ambiguous, as in~\figref{cw-force}.  For two purely
chiral materials in the EMA regime, the zero-frequency component of
the force difference is exactly zero~\cite{Zhao09}, so the magnitude
of the zero-frequency component is an indication of whether the force
difference arises from chirality alone.
In~\figref{omega-results}(Inset) we display the frequency integrand at
$z=3.6a$, indicating that the force difference in the isotropic case
is indeed due to chirality.  However, we have found that the frequency
integrand for less isotropic unit cells has a large zero-frequency
component, indicating that anisotropy and proximity effects play a
large role in that case.  This could also explain the results
of~\figref{cw-force} for larger $z$.  In conclusion, to observe an
unambiguous effect of chirality in isotropic, chiral metamaterials,
one must measure the total force to 4 digits of accuracy. Given that
$a \gg 1~\mu$m for realistic systems, and that anisotropies in the
structure can result in much stronger force variations, this effect is
unlikely to be relevant for applications.  This does not preclude the
possibility of other interesting Casimir effects in metamaterials
(especially if they rely on the $\omega \rightarrow 0$ response of the
metamaterial), but it imposes serious limitations for finite-$\omega$
effects.

Finally, we comment on the validity of the effective medium picture
and the dilute approximation.  One can compute the Casimir force and
the force difference from the EMA parameters
of~\figref{omega-medium}. For the range of $z/a$
in~\figref{omega-results}, this gives the correct force between the
two media but overestimates the force difference by roughly 30
percent.  This is because the error terms in the EMA above are
$O(k_\perp^4)$, the same order as the chirality contributions $\sim
R_\mathrm{sp}^2$ to the force~\cite{Zhao09}.  The Clausius--Mossotti
(C--M) equation allows us to check the validity of the dilute
approximation.  C--M relates polarizabilities of particles to the
effective-medium parameters~\cite{Collin91,Tretyakov2005}.  We compute
the polarizabilities from C--M, working backwards from the
effective-medium parameters of~\figref{omega-medium} (expanding C--M
to first order in the polarizability), and plug this into the full
C--M to compute the non-dilute correction.  The relative force
difference of~\figref{omega-results} is changed by $< 15\%$.  C--M is
obtained in the static limit, whereas at finite imaginary $\omega$ the
exponential field decay reduces the interactions, so this is an
overestimate.

We thank P. Bermel and S. J. Rahi for useful discussions.  This work
was supported by the Army Research Office through the ISN under
Contract No. W911NF-07-D-0004, and by DARPA under contract
N66001-09-1-2070-DOD and under DOE/NNSA contract DE-AC52-06NA25396.


\end{document}